\documentclass[a4paper]{article}

\usepackage{INTERSPEECH2022}
\usepackage{url}
\usepackage{enumerate}
\usepackage{multirow}
\usepackage{booktabs}
\usepackage{makecell}
\usepackage{threeparttable}

\title{SpeechFormer: A Hierarchical Efficient Framework \\Incorporating the Characteristics of Speech}

\name{Weidong Chen$^1$, Xiaofen Xing$^1$, Xiangmin Xu$^1$\sthanks{Corresponding author}, Jianxin Pang$^2$, Lan Du$^3$}
\address{$^1$School of Electronic and Information Engineering, South China University of Technology, China\\ $^2$UBTECH Robotics Corp, China  \qquad  $^3$iFLYTEK Research, China}
\email{eewdchen@mail.scut.edu.cn,\{xfxing,xmxu\}@scut.edu.cn,walton@ubtrobot.com,landu@iflytek.com}

\begin{document}

\maketitle
\begin{abstract}
Transformer has obtained promising results on cognitive speech signal processing field, which is of interest in various applications ranging from emotion to neurocognitive disorder analysis.
However, most works treat speech signal as a whole, leading to the neglect of the pronunciation structure that is unique to speech and reflects the cognitive process. 
Meanwhile, Transformer has heavy computational burden due to its full attention operation.
In this paper, a hierarchical efficient framework, called SpeechFormer, which considers the structural characteristics of speech, is proposed and can be served as a general-purpose backbone for cognitive speech signal processing.
The proposed SpeechFormer consists of frame, phoneme, word and utterance stages in succession, each performing a neighboring attention according to the structural pattern of speech with high computational efficiency.
SpeechFormer is evaluated on speech emotion recognition (IEMOCAP \& MELD) and neurocognitive disorder detection (Pitt \& DAIC-WOZ) tasks,
and the results show that SpeechFormer outperforms the standard Transformer-based framework while greatly reducing the computational cost.
Furthermore, our SpeechFormer achieves comparable results to the state-of-the-art approaches.
\end{abstract}

\section{Introduction}
Speech signal is able to express the most information in the simplest way \cite{perception}. 
Speech based emotion and neurocognitive disorder analysis, which is collectively referred to as \textbf{co}gnitive \textbf{s}peech \textbf{s}ignal \textbf{p}rocessing (CoSSP), 
has covered a wide area of applications,
including speech emotion recognition (SER), depression classification, Alzheimer's disease (AD) detection and so on. Because of its broad application value, CoSSP is becoming an increasing interest in speech signal processing field.

In the last century, Hidden Markov model, which is a statistical Markov model and assumes the system being modeled to be a Markov process, is proposed to model speech signal \cite{HMM1,HMM3}. After that, more and more machine learning methods, such as decision tree \cite{Tree1,Tree2} and restricted Boltzmann machine \cite{Boltzmann1,Boltzmann12} and so on, are applied to CoSSP. Recently, with the development of deep learning, Convolutional Neural Network \cite{iemocap_2, depression_cnn1, pitt_3, daiz_1}, Recurrent Neural Network and its variants \cite{SER_RNN1, SER_RNN2, SER_RNN3, daiz_2} are proposed and achieve promising results. 
After that, deep learning methods deliver superior performances in CoSSP filed.

Inspired by the global attention mechanism, Transformer \cite{transformer}, which is outstanding in modeling long-range dependencies in the sequence, has achieved great success in natural language processing (NLP). Although the original Transformer is designed for machine translation task in NLP, researchers are active in investigating its adaptation to many other fields, specifically computer vision \cite{ViT, Swin}. 
Certainly, several attempts also have been made in CoSSP field \cite{speech_use_trans2, meld_c1, meld_c3, ksT}.

However, most studies omit the natural characteristics of speech while using the attention with Transformer, leading to the neglect of the pronunciation structure that is unique to speech and conveys lots of messages. 
For example, even if I don't understand Greek, I can still determine the emotion in a Greek recording by utilizing the characteristics in speech, such as articulation, prolongation and the dynamic change of speech sound.
Meanwhile, Transformer has heavy computational burden, as the computational complexity of its full attention is quadratic to input length.
In other words, the standard Transformer needs to incorporate the characteristics of speech before all its performance can be exploited in CoSSP field.



\begin{figure}[t]
\centering
\includegraphics[width=\linewidth]{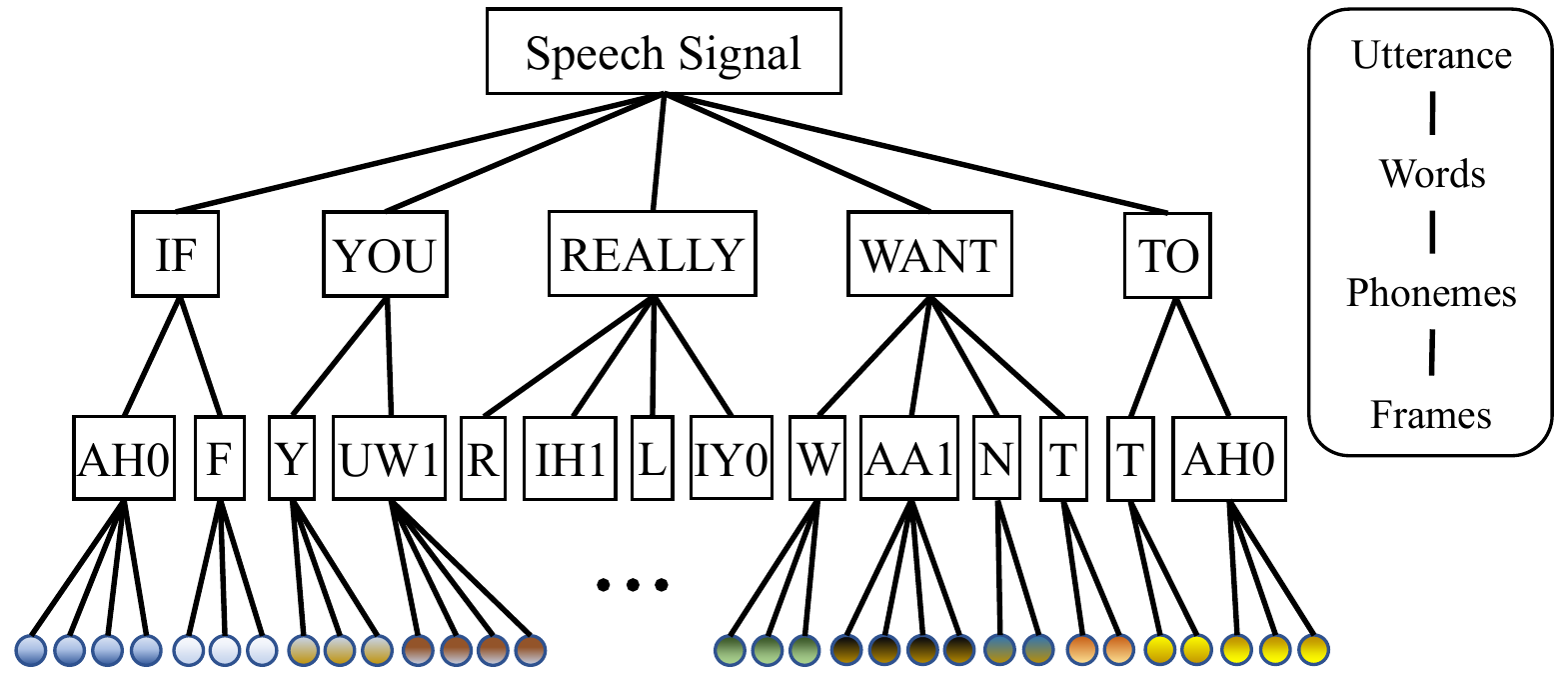}
\caption{The natural structure of a speech signal, in which several frames constitute a phoneme, several phonemes constitute a word, and multiple words form an utterance in speech signal.
}
\label{fig_1}
\end{figure}

\begin{figure*}[t]
\centering
\includegraphics[width=\linewidth]{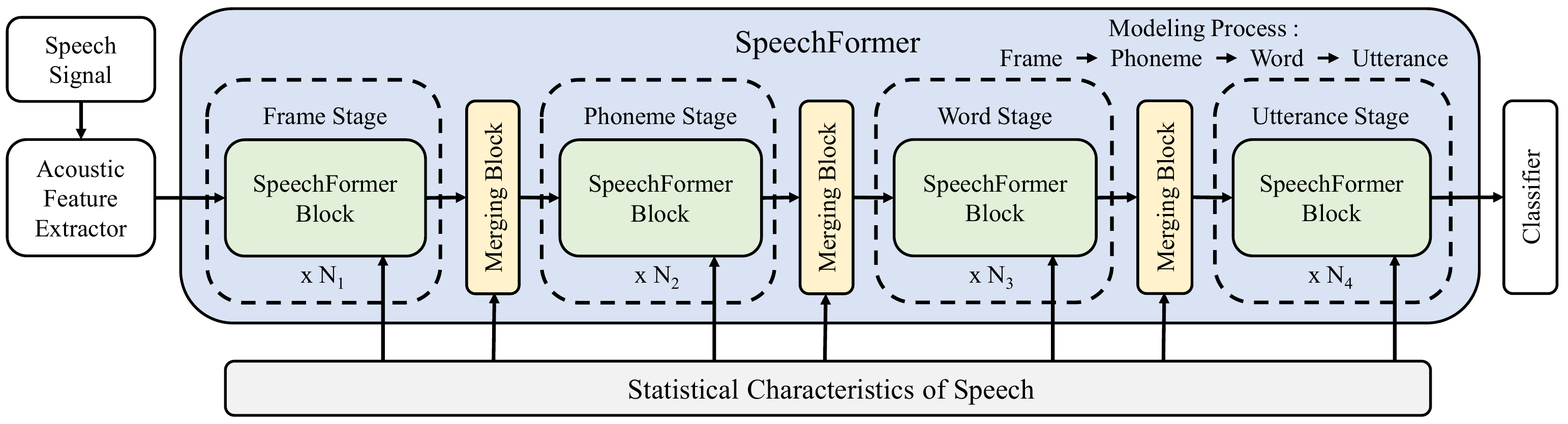}
\caption{Overview of the proposed SpeechFormer framework.
}
\label{fig_2}
\end{figure*}

To solve the above problems, we should rethink the structure of speech signal first. 
As shown in Figure~\ref{fig_1}, we can observe that an utterance consists of several words, a word consists of several phonemes (e.g. word `WANT' includes phonemes `W', `AA1', `N' and `T') and a phoneme consists of several frames (e.g. phoneme `AA1' includes four frames). 
This progressive structure reveals the importance of the interaction between adjacent elements and indicates that we can model the speech signal hierarchically 
based on the nature of pronunciation.
Therefore, a hierarchical framework, called SpeechFormer, which consists of frame, phoneme, word and utterance stages, is proposed to model the signal step-by-step. 
Firstly, we capture the frame, phoneme and word-level features through the first three stages and merge the features between two successive stages, both of which are performed under the instruction of the characteristics of speech with high computational efficiency.
At last, an utterance stage is applied to gather all the word-level features and generate an utterance-level representation for classification.
The contributions of this paper can be summarized as follows:
\begin{itemize}
\item We propose a hierarchical efficient framework, called SpeechFormer, to serve as a general-purpose backbone for cognitive speech signal processing. SpeechFormer improves the modeling process by incorporating the characteristics of speech, which follows the natural pronunciation structure of the input speech signal.
\item We evaluate SpeechFormer on IEMOCAP \cite{IEMOCAP}, MELD \cite{meld}, Pitt \cite{pitt} and DAIC-WOZ \cite{avec}, and demonstrate that SpeechFormer substantially outperforms the vanilla Transformer-based framework  in terms of performance and computational efficiency. Moreovre, SpeechFormer achieves comparable results to the state-of-the-art approaches. Our codes are publicly available at \url{https://github.com/HappyColor/SpeechFormer}.
\end{itemize}



\section{Methodology}
The proposed SpeechFormer, as shown in Figure~\ref{fig_2}, mainly consists of four stages and three merging blocks (M-Blocks). In which, frame stage (F-Stage), phoneme stage (P-Stage) and word stage (W-Stage) are used to learn features of different levels, and utterance stage (U-Stage) aims to
generate a global representation for classification. Three M-Blocks refine the redundant features between two consecutive stages by reducing the number of tokens. Moreover, an additional branch is employed to provide the statistical natures of speech signal.
More details will be introduced in the following subsections.

\subsection{Vanilla Transformer}
\begin{figure}[t]
\centering
\includegraphics[width=0.7\linewidth]{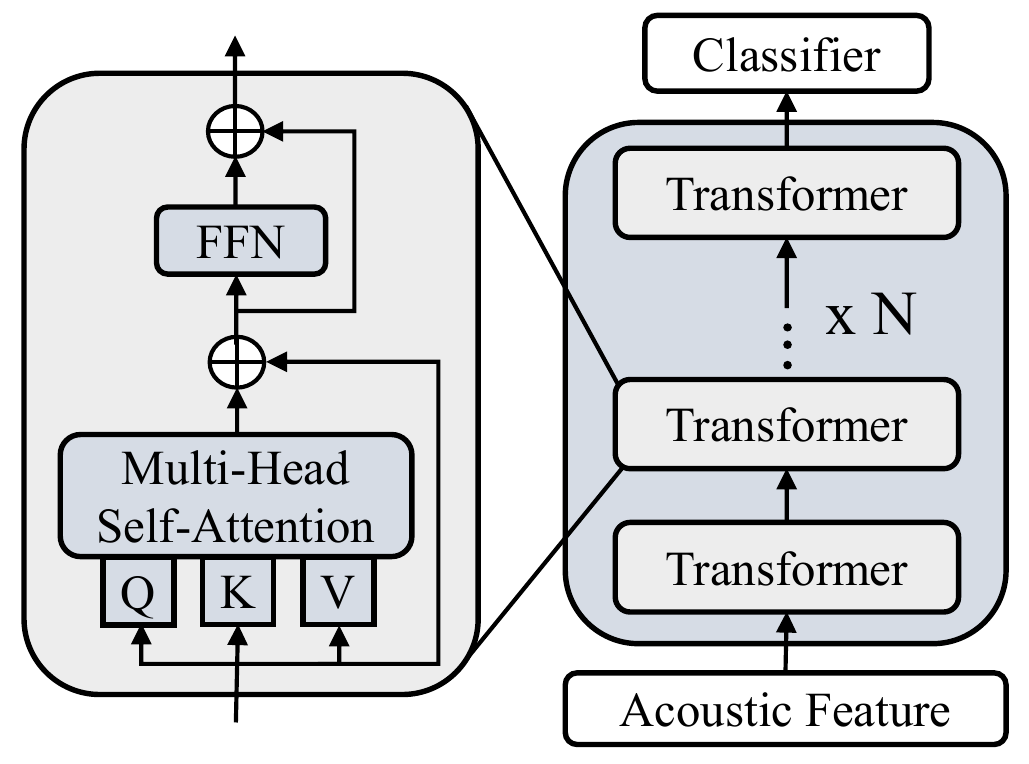}
\caption{The structure of the vanilla Transformer (left) and the baseline framework used in this paper (right). Positional encoding and layer normalization \cite{LN} haven't plotted for brevity.
}
\label{fig_3}
\end{figure}

Transformer (refers only to its encoder part in this paper), as shown in the left part of Figure~\ref{fig_3}, consists of two sub-layers of Multi-Head Self-Attention (MSA) and Feed-Forward Network (FFN). 
MSA is at the core of Transformer and we will give a brief introduction to it. More details can be found in \cite{transformer}.

\subsubsection{MSA in vanilla Transformer}
For a sequential input $x\in\mathbb{R}^{T \times d_{m}}$, where $\it T$ and $\it d_{m}$ are the length and dimension of input, respectively. Transformer first obtains query \textbf{Q}, key \textbf{K} and value \textbf{V} by applying three projections to $\it x$. \textbf{Q}\textbf{K}\textbf{V} are further divided into $\it h$ parts, 
producing $\it d_{h}$ dimensional features,
where $d_{h}=d_{m}/h$ and $\it h$ is the number of heads. 
Each head performs Single-Head Self-Attention (SSA) and the output value of each head is concatenated to form the final output of MSA.
SSA is depicted as fellows:

\begin{equation}
  \rm SSA(\textbf{Q},\textbf{K},\textbf{V}) = Softmax(\frac{{\textbf{QK}}^{T}}{\it {d_{h}}}) {\textbf{V}}
  \label{eq1}
\end{equation}

\subsection{SpeechFormer framework}
The vanilla Transformer-based framework, illustrated in the right part of Figure~\ref{fig_3}, neglects the implicit relations in speech. 
Conversely,  our SpeechFormer, shown in Figure~\ref{fig_2}, makes use of these relations to model speech signal in a hierarchical manner. To be specific, 
SpeechFormer block employs Speech-based Multi-Head Self-Attention (Speech-MSA) to capture the relations between adjacent elements.
Merging block is applied to refine the redundant feature. Both of them are performed under the instruction of the statistical characteristics of speech.

\subsubsection{Statistical characteristics of speech}

\textbf{Elements of speech signal.} Phoneme is the minimal sound unit in language, recycled to form all our spoken words. Multiple words are arranged together to form an utterance that is recorded in a wave file. In data format terms, the digital speech signal is divided into numerous frames. Each frame contains information at that particular point in time. Therefore, frame is the basic processing unit in digital system, which then gradually forms a phoneme, a word, and finally an utterance.

\noindent \textbf{Time duration.} The frame length is literally the size of the window during, which can be set manually. Phonemes are of different lengths, varying from 50 ms to 200 ms. 
To analyse the duration of word, we use P2FA \cite{P2FA} to extract the phonemes from corpora used and find that more than 90\% of words contain less than 5 phonemes. 
Thus, the duration of word is considered as 250 ms to 1000 ms (5$\times$ the duration of phoneme). Time duration divided by $hop$ is the approximate number of tokens in the feature, where $hop$ indicates the feature's hop length.

\begin{figure}[t]
\centering
\includegraphics[width=0.7\linewidth]{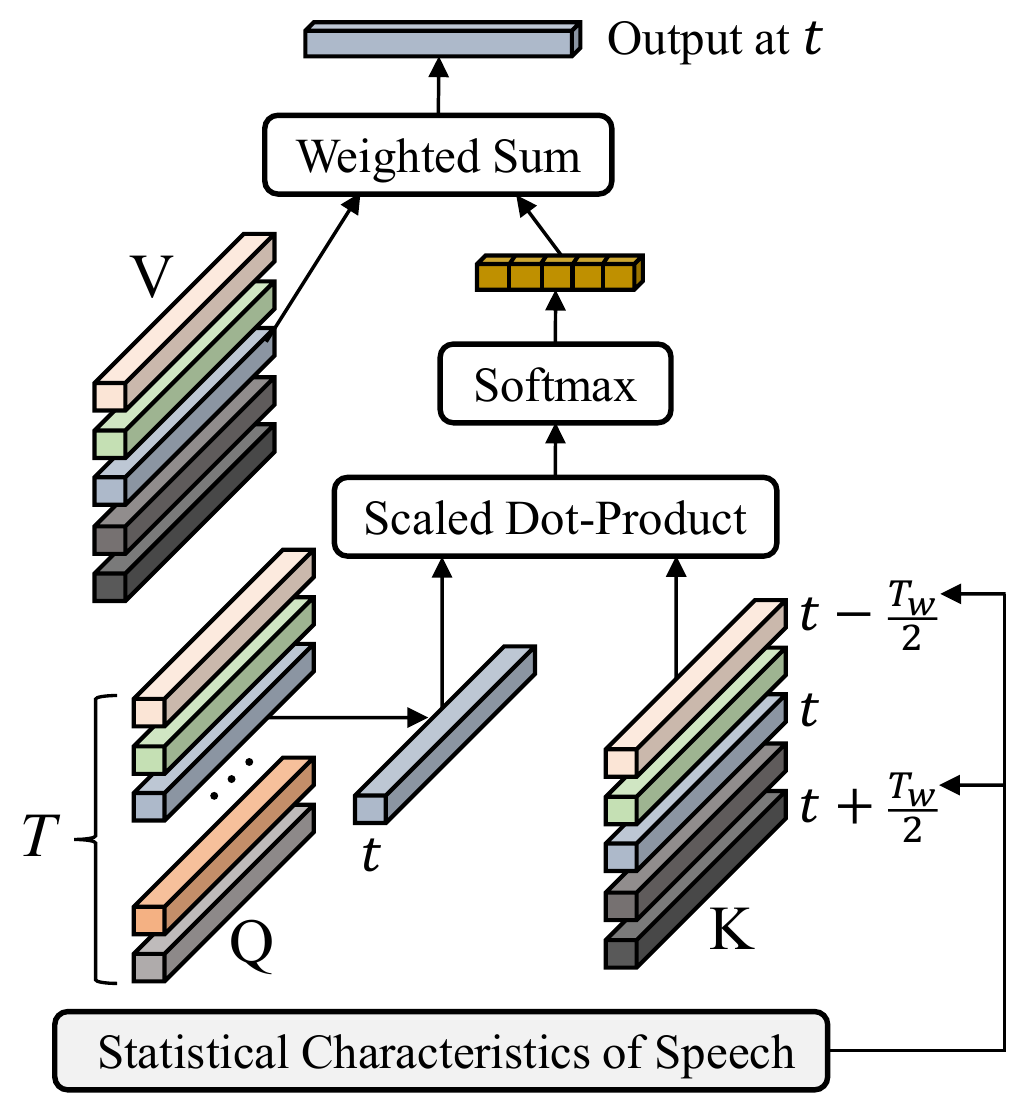}
\caption{Speech-based Multi-Head Self-Attention (Speech-MSA) at time step $t$. $T_{w}$ is a statistical value based on speech characteristics. Head number here is set to 1 for illustration.
}
\label{fig_4}
\end{figure}

\subsubsection{Speech-MSA in SpeechFormer block}
The only difference between SpeechFormer block and standard Transformer is the replacement of MSA with Speech-MSA. 
As shown in Figure~\ref{fig_4}, Speech-MSA applies a window $T_{w}$ to limit the full attention computation to a small scope of adjacent tokens, which can greatly relieve the computational burden.
Furthermore, as displayed in Table~\ref{tab_1}, the value of $T_{w}$ ensures that the Speech-MSA in the first three stages can learn the interactions between neighboring frames, phonemes and words, respectively. And in U-Stage, $T_{w}$ is set to the length of its input such that a global representation is learnt.
The complete output of Speech-MSA is calculated by following steps:

\begin{enumerate}[(i)]
\item Apply for a window $T_{w}$ according to the current stage-level. The value and explanation are shown in Table~\ref{tab_1}.
\item Select the $(t-\frac{{T_{w}}}{{2}})$-th to $(t+\frac{{T_{w}}}{{2}})$-th tokens in \textbf{K}, each performing a scaled dot-product with the $t$-th token in \textbf{Q} to produce a score. All the scores are concatenated and scaled by Softmax to generate the attention weights.
\item Fetch the $(t-\frac{{T_{w}}}{{2}})$-th to $(t+\frac{{T_{w}}}{{2}})$-th tokens in \textbf{V} and calculate the sum of each token multiplied by its weight. The result is the output of the $t$-th token in Speech-MSA.
\item Return to (ii) until the value of $t$ varies from 1 to $T$.
\end{enumerate}


\subsubsection{Merging block in SpeechFormer framework}
The merging blocks between each two successive stages refine features under the instruction of the characteristics of speech, making a hierarchical framework which follows the natural structure of speech signal. Each merging block consists of an average pooling layer and a linear layer in succession. The first layer merges $M$ consecutive tokens in the input, where $M$ is the merging scale, and the second layer expands the embedding dimension of the input by a factor of $r$. Each merging block aims to prepare the appropriate feature for its following stage as input. 
Specifically, each token should represent a sub-phoneme in P-Stage such that P-Stage can model the interaction between neighboring phonemes. 
Therefore, the $M_1$ before P-Stage should be no less than the minimum length of phoneme. Similarly, the $M_2$ before W-Stage should be no less than the minimum length of word. More details are displayed in Table~\ref{tab_1}.

\begin{table}[t]
    \caption{The values of window $T_{w}$ and merging scale $M$. Statistical duration is used instead of the actual duration.}
    \label{tab_1}
    \centering
    \begin{tabular}{|c|c|l|}
    \hline
    Module      & $T_{w}$ or $M$            &  Description                       \\ \hline
    F-Stage     & 50 ms / $hop_1$           &  Min. length of phoneme            \\ \hline
    M-Block     & 50 ms / $hop_1$           &  Min. length of phoneme            \\ \hline
    P-Stage     & 400 ms / $hop_2$          &  $2 \times$Max. length of phoneme  \\ \hline
    M-Block     & 250 ms / $hop_2$          &  Min. length of word               \\ \hline
    W-Stage     & 2000 ms / $hop_3$         &  $2 \times$Max. length of word     \\ \hline
    M-Block     & 1000 ms / $hop_3$         &  Max. length of word               \\ \hline
    U-Stage     & $T$                       &  Input length                      \\ \hline
    \multicolumn{3}{|l|}{Notes: $hop_2 = M_{1}hop_{1}$, $hop_3 = M_{2}hop_{2}$}   \\ \hline
    \end{tabular}
\end{table}

\section{EXPERIMENTS}

\subsection{Datasets}

\noindent \textbf{IEMOCAP} is the most widely used dataset in SER field. 
We select 5,531 utterances from four emotion categories: angry, neutral, happy (\& excited) and sad.
Experiments are conducted in leave-one-session-out cross-validation strategy.

\noindent \textbf{MELD} is another dataset used in SER. It consists of 13,708 utterances with seven emotions. The dataset is split into train, validation and test sets, and the scores on test set are reported.

\noindent \textbf{Pitt} corpus is used in AD detection field.
The AD patients and healthy controls are asked to take the ``Cookie Theft" picture description task \cite{cookie} to produce recordings. 
Experiments are conducted in speaker-independent 10-fold cross-validation strategy.

\noindent \textbf{DAIC-WOZ} corpus
, used in AVEC 2016 \cite{avec2016}, 
contains recordings labeled depressed / not depressed.
Since the labels of test data are not provided, we report the results on validation set.

\subsection{Experimental setup}

\noindent \textbf{Acoustic features.} Three types of acoustic features, named spectrogram (Spec), Log-Mel spectrogram (Logmel) and pre-trained Wav2vec \cite{wav2vec}, are extracted in this paper. The window sizes of Spec and Logmel are set to 20 and 25 ms, respectively. The number of Mel frequency bands used when extracting Logmel is 128 (in IEMOCAP \& DAIC-WOZ) and 256 (in MELD \& Pitt). The original hop length $hop_1$ is set to 10 ms by default.

\begin{table}[t]
    \caption{The training settings on four corpora. The learning rate drops to 1\% of the original gradually by cosine annealing. WA: weighted accuracy, UA: unweighted arruracy, WF1: weighted average F1, MF1: macro average F1.}
    \label{tab_2}
    \centering
    \begin{threeparttable}
    \begin{tabular}{|c|c|c|c|c|}
    \hline
    Dataset      & Epoch     & Batch    & LR      & Criterion         \\ \hline
    IEMOCAP      & 120       & 32       & 0.0005  & WA \& UA          \\ \hline
    MELD         & 120       & 32       & 0.001   & WF1               \\ \hline
    Pitt\tnote{\#}         & 80        & 32       & 0.0005  & WA \& UA          \\ \hline
    DAIC-WOZ\tnote{\#}     & 60        & 16       & 0.0005  & MF1               \\ \hline
    \end{tabular}
    \begin{tablenotes}
        \footnotesize
        \item[\#] Segment-level samples are used for training, and the subject-level result obtained by majority vote is used for evaluation.
    \end{tablenotes}
    \end{threeparttable}
\end{table}

\makeatletter
\def\hlinew#1{%
  \noalign{\ifnum0=`}\fi\hrule \@height #1 \futurelet
   \reserved@a\@xhline}
\makeatother

\begin{table*}[t]
    \caption{The performances of the baseline framework and the proposed SpeechFormer framework on IEMOCAP, MELD, Pitt and DAIC-WOZ corpora. Model size (Params) and theoretical computational complexity (FLOPs) are also listed for comparison.}
    \label{tab_3}
    \centering
    \begin{tabular}{|c|c|c|c|c|c|c|c|c|c|c|}
    \hlinew{1.0pt}
    \multirow{2}{*}{Feature} & \multirow{2}{*}{Architecture}  &  \multicolumn{5}{c|}{Speech emotion recognition: IEMOCAP}  &  \multicolumn{4}{c|}{Speech emotion recognition: MELD} \\ \cline{3-11} 
                             &                & Input size                      & Params  & FLOPs  & WA    & UA    & Input size  &  Params  & FLOPs  & WF1    \\ \hline
    \multirow{3}{*}{Spec}    & Baseline       & \multirow{3}{*}{653$\times$161} & 1.56M   & 2.64G  & 0.572 & 0.585 & \multirow{3}{*}{449$\times$442} & 11.77M & 7.35G & 0.368\\
                             & SpeechFormer-S &                                 & 1.64M   & \textbf{0.23G}  & 0.572 & 0.588 &                        & 12.35M & \textbf{1.16G} & \textbf{0.382}\\
                             & SpeechFormer-B &                                 & 3.25M   & 0.24G  & \textbf{0.580} & \textbf{0.594} &               & 24.53M & 1.22G & 0.378\\ \hline
    \multirow{3}{*}{Logmel}  & Baseline       & \multirow{3}{*}{651$\times$128} & 1.00M   & 1.94G  & 0.570 & 0.584 & \multirow{3}{*}{446$\times$256} & 3.99M  & 2.98G & 0.389\\
                             & SpeechFormer-S &                                 & 1.05M   & \textbf{0.14G}  & 0.578 & 0.596 &                        & 4.19M  & \textbf{0.39G} & 0.389\\
                             & SpeechFormer-B &                                 & 2.09M   & 0.15G  & \textbf{0.582} & \textbf{0.598} &               & 8.32M  & 0.41G & \textbf{0.396}\\\hline
    \multirow{3}{*}{Wav2vec} & Baseline       & \multirow{3}{*}{651$\times$512} & 15.93M  & 15.45G & 0.618 & 0.629 & \multirow{3}{*}{446$\times$512} & 15.93M & 9.46G & 0.409\\
                             & SpeechFormer-S &                                 & 16.72M  & \textbf{2.28G}  & \textbf{0.629} & \textbf{0.645} &      & 16.72M & \textbf{1.56G} & 0.418\\
                             & SpeechFormer-B &                                 & 33.21M  & 2.40G  & 0.623 & 0.636 &                                 & 33.21M & 1.64G & \textbf{0.419}\\\hlinew{1.0pt}
    \multirow{2}{*}{Feature} & \multirow{2}{*}{Architecture}  &  \multicolumn{5}{c|}{Alzheimer’s disease detection: Pitt}  &  \multicolumn{4}{c|}{Depression classification: DAIC-WOZ} \\ \cline{3-11} 
                             &                & Input size                      & Params  & FLOPs  & WA    & UA    & Input size  &  Params  & FLOPs & MF1    \\ \hline
    \multirow{3}{*}{Spec}    & Baseline       & \multirow{3}{*}{656$\times$442} & 11.77M  & 12.17G & 0.692 & 0.678 & \multirow{3}{*}{1076$\times$161} & 1.56M  & 6.10G  &0.530\\
                             & SpeechFormer-S &                                 & 12.35M  & \textbf{1.70G}  & \textbf{0.698} & \textbf{0.695} &       & 1.64M  & \textbf{0.37G}  &0.551\\
                             & SpeechFormer-B &                                 & 24.53M  & 1.78G  & 0.694 & 0.686 &                                  & 3.25M  & 0.39G  &\textbf{0.558}\\ \hline
    \multirow{3}{*}{Logmel}  & Baseline       & \multirow{3}{*}{658$\times$256} & 3.99M   & 5.25G  & \textbf{0.680} & 0.656 & \multirow{3}{*}{1074$\times$128} & 1.00M  & 4.60G  &0.583\\
                             & SpeechFormer-S &                                 & 4.19M   & \textbf{0.58G}  & 0.667 & 0.645 &                         & 1.05M  & \textbf{0.24G}  &\textbf{0.627}\\
                             & SpeechFormer-B &                                 & 8.31M   & 0.61G  & 0.675 & \textbf{0.669} &                         & 2.09M  & 0.25G  &0.608\\\hline
    \multirow{3}{*}{Wav2vec} & Baseline       & \multirow{3}{*}{658$\times$512} & 15.93M  & 15.67G & 0.751 & 0.746 & \multirow{3}{*}{1074$\times$512} & 15.93M & 31.07G &0.657\\
                             & SpeechFormer-S &                                 & 16.72M  & \textbf{2.30G}  & 0.752 & 0.751 &                         & 16.72M & \textbf{3.75G}  &0.676\\
                             & SpeechFormer-B &                                 & 33.21M  & 2.42G  & \textbf{0.757} & \textbf{0.752} &                & 33.21M & 3.93G  &\textbf{0.694}\\\hlinew{1.0pt}
    \end{tabular}
\end{table*}

\noindent \textbf{Hyper-parameters.} The hyper-parameters used in different datasets for training are listed in Table~\ref{tab_2}. SGD \cite{sgd} is employed to optimize the model. The number of Transformers $\rm N$ used in baseline framework is 12. 
The number of heads used in multi-head attention is 8.
As for the SpeechFormer framework, we introduce two versions with different model sizes and computational complexity. Their hyper-parameters are:

\begin{itemize}
\item SpeechFormer-S: $\rm N_1$$ \sim $$\rm N_4$ = \{2, 2, 4, 4\}, $r$ = \{1, 1, 1\} 
\item SpeechFormer-B: $\rm N_1$$ \sim $$\rm N_4$ = \{2, 2, 4, 4\}, $r$ = \{1, 1, 2\}
\end{itemize}


\subsection{Experimental results and analysis}
\subsubsection{Comparison to the baseline framework}
The comparison results of the proposed SpeechFormer and the baseline framework on IEMOCAP, MELD, Pitt and DAIC-WOZ are summarized in Table~\ref{tab_3}. 
Speech emotion recognition, Alzheimer’s disease detection and depression classification tasks are involved. 
From Table~\ref{tab_3}, we can observe that the model sizes of SpeechFormer-S and baseline are very close, where the former is slightly larger than the later. 
However, the theoretical computational complexity\footnote{We omit Softmax computation in determining complexity.} (FLOPs) of SpeechFormer-S on four corpora 
are about 10.2\%, 15.1\%, 13.2\% and 7.8\% of the baseline, respectively.
SpeechFormer-B expands the model size of SpeechFormer-S by approximately two times, while keeping the FLOPs comparable. While conducting \textbf{speech emotion recognition}, the baseline results are lower than those of SpeechFormer, regardless of the feature used. 
When performing \textbf{Alzheimer’s disease detection}, the baseline achieves better WA using Logmel as input. In other cases, SpeechFormer obtains higher performances. As for \textbf{depression classification}, SpeechFormer framework substantially outperforms the baseline in all cases. In summary, our SpeechFormer framework generally outperforms the vanilla Transformer-based framework in terms of performance and computational efficiency.


\subsubsection{Comparison to previous state-of-the-art}
Table~\ref{tab_4} gives the comparison among SpeechFormer with some known systems on four corpora, in which, all systems utilize only speech feature as input to allow a fair comparison. 
On IEMOCAP, SpeechFormer-S achieves comparable performances to \cite{iemocap_4}: +0.6\% WA.
When evaluated on MELD, Pitt and DAIC-WOZ, SpeechFormer-B outperforms other comparisons with promising gains: +1.7\% WF1 over \cite{meld_c1}, +1.8\% (+11.1\%) WA (UA) over \cite{pitt_2} and +3.4\% MF1 over \cite{daiz_2}, respectively.

\begin{table}[t]
    \caption{The performances of the state-of-the-art approaches and the proposed SpeechFormer on four corpora.}
    \label{tab_4}
    \centering
    \begin{tabular}{|c|c|c|c|}
    \hline
    Dataset                  & Method         & WA    & UA      \\ \hline
    \multirow{4}{*}{IEMOCAP} & [Guo \emph{et al.},2021]\cite{iemocap_2} & 0.613 & 0.604   \\ 
                             & [Yin \emph{et al.},2021]\cite{iemocap_4} & 0.623 & -   \\ \cline{2-2}
                             & SpeechFormer-S & \textbf{0.629} & \textbf{0.645}   \\ 
                             & SpeechFormer-B & 0.623 & 0.636   \\ \hline
    \multirow{4}{*}{MELD}    & [Liang \emph{et al.},2020]\cite{meld_c1}     &\multicolumn{2}{|c|}{0.402 (WF1)}\\ 
                             & [Lian \emph{et al.},2021]\cite{meld_c3}      &\multicolumn{2}{|c|}{0.382 (WF1)}\\ \cline{2-2}
                             & SpeechFormer-S                               &\multicolumn{2}{|c|}{0.418 (WF1)}\\ 
                             & SpeechFormer-B                               &\multicolumn{2}{|c|}{\textbf{0.419} (WF1)}  \\\hline
    \multirow{4}{*}{Pitt}    & [Makiuchi \emph{et al.},2021]\cite{pitt_3}   & 0.731 & -      \\ 
                             & [Bertini \emph{et al.},2022]\cite{pitt_2}    & 0.739 & 0.641  \\ \cline{2-2}
                             & SpeechFormer-S                               & 0.752 & 0.751  \\ 
                             & SpeechFormer-B       & \textbf{0.757}        & \textbf{0.752}   \\ \hline
    \multirow{4}{*}{DAIC-WOZ}& [Solieman \emph{et al.},2021]\cite{daiz_1}   &\multicolumn{2}{|c|}{0.610 (MF1)}\\ 
                             & [Dumpala \emph{et al.},2021]\cite{daiz_2}    &\multicolumn{2}{|c|}{0.660 (MF1)}\\ \cline{2-2}
                             & SpeechFormer-S                               &\multicolumn{2}{|c|}{0.676 (MF1)}\\ 
                             & SpeechFormer-B                               &\multicolumn{2}{|c|}{\textbf{0.694} (MF1)}\\ \hline
    \end{tabular}
\end{table}

\section{Conclusion}
A hierarchical efficient framework, named SpeechFormer, has been proposed for cognitive speech signal processing. 
SpeechFormer applies four stages to learn the representation, following the natural structure of speech. The Speech-MSA and merging block in SpeechFormer model speech signal efficiently by incorporating the statistical characteristics of speech.
Experimental results on four corpora, including three speech-related cognitive tasks, demonstrate that SpeechFormer outperforms the vanilla Transformer-based framework in terms of performance and computational efficiency. The comparison to state-of-the-art also verifies the effectiveness of SpeechFormer. 
In the future, we plan to extend SpeechFormer to other speech tasks, such as automatic speech recognition and speaker identification.

\bibliographystyle{IEEEtran}

\bibliography{mybib}

\begin{thebibliography}{10}
\providecommand{\url}[1]{#1}
\csname url@samestyle\endcsname
\providecommand{\newblock}{\relax}
\providecommand{\bibinfo}[2]{#2}
\providecommand{\BIBentrySTDinterwordspacing}{\spaceskip=0pt\relax}
\providecommand{\BIBentryALTinterwordstretchfactor}{4}
\providecommand{\BIBentryALTinterwordspacing}{\spaceskip=\fontdimen2\font plus
\BIBentryALTinterwordstretchfactor\fontdimen3\font minus
  \fontdimen4\font\relax}
\providecommand{\BIBforeignlanguage}[2]{{%
\expandafter\ifx\csname l@#1\endcsname\relax
\typeout{** WARNING: IEEEtran.bst: No hyphenation pattern has been}%
\typeout{** loaded for the language `#1'. Using the pattern for}%
\typeout{** the default language instead.}%
\else
\language=\csname l@#1\endcsname
\fi
#2}}
\providecommand{\BIBdecl}{\relax}
\BIBdecl

\bibitem{perception}
B.~Moore, L.~Tyler, and W.~Marslen-Wilson, ``Introduction. the perception of
  speech: from sound to meaning,'' \emph{Philosophical transactions of the
  Royal Society of London. Series B, Biological sciences}, vol. 363, no. 1493,
  pp. 917--921, Mar. 2008.

\bibitem{HMM1}
K.~Tokuda, T.~Kobayashi, and S.~Imai, ``{Speech parameter generation from HMM
  using dynamic features},'' in \emph{IEEE International Conference on
  Acoustics, Speech and Signal Processing (ICASSP)}, 1995, pp. 660--663.

\bibitem{HMM3}
B.~Schuller, G.~Rigoll, and M.~Lang, ``{Hidden Markov model-based speech
  emotion recognition},'' in \emph{IEEE International Conference on Acoustics,
  Speech, and Signal Processing (ICASSP)}, 2003, pp. II--1.

\bibitem{Tree1}
W.~Reichl and W.~Chou, ``Robust decision tree state tying for continuous speech
  recognition,'' \emph{IEEE Transactions on Speech and Audio Processing},
  vol.~8, no.~5, pp. 555--566, 2000.

\bibitem{Tree2}
L.~Yang, D.~Jiang, L.~He, E.~Pei, M.~C. Oveneke, and H.~Sahli, ``Decision tree
  based depression classification from audio video and language information,''
  in \emph{Proceedings of the 6th International Workshop on Audio/Visual
  Emotion Challenge}, 2016, pp. 89--96.

\bibitem{Boltzmann1}
A.~Stuhlsatz, C.~Meyer, F.~Eyben, T.~Zielke, G.~Meier, and B.~Schuller, ``Deep
  neural networks for acoustic emotion recognition: Raising the benchmarks,''
  in \emph{IEEE International Conference on Acoustics, Speech and Signal
  Processing (ICASSP)}, 2011, pp. 5688--5691.

\bibitem{Boltzmann12}
K.~Poon-Feng, D.-Y. Huang, M.~Dong, and H.~Li, ``Acoustic emotion recognition
  based on fusion of multiple feature-dependent deep boltzmann machines,'' in
  \emph{The 9th International Symposium on Chinese Spoken Language Processing},
  2014, pp. 584--588.

\bibitem{iemocap_2}
L.~Guo, L.~Wang, C.~Xu, J.~Dang, E.~S. Chng, and H.~Li, ``Representation
  learning with spectro-temporal-channel attention for speech emotion
  recognition,'' in \emph{IEEE International Conference on Acoustics, Speech
  and Signal Processing (ICASSP)}, 2021, pp. 6304--6308.

\bibitem{depression_cnn1}
N.~Seneviratne and C.~Espy-Wilson, ``Generalized dilated {CNN} models for
  depression detection using inverted vocal tract variables,'' in \emph{Proc.
  Interspeech}, 2021, pp. 4513--4517.

\bibitem{pitt_3}
M.~R. Makiuchi, T.~Warnita, N.~Inoue, K.~Shinoda, M.~Yoshimura, M.~Kitazawa,
  K.~Funaki, Y.~Eguchi, and T.~Kishimoto, ``Speech paralinguistic approach for
  detecting dementia using gated convolutional neural network,'' \emph{IEICE
  transactions on Information and Systems}, vol. 104, no.~11, pp. 1930--1940,
  2021.

\bibitem{daiz_1}
H.~Solieman and E.~A. Pustozerov, ``The detection of depression using
  multimodal models based on text and voice quality features,'' in \emph{IEEE
  Conference of Russian Young Researchers in Electrical and Electronic
  Engineering}, 2021, pp. 1843--1848.

\bibitem{SER_RNN1}
J.~Lee and I.~Tashev, ``High-level feature representation using recurrent
  neural network for speech emotion recognition,'' in \emph{Proc. Interspeech},
  2015, pp. 1537--1540.

\bibitem{SER_RNN2}
S.~Mirsamadi, E.~Barsoum, and C.~Zhang, ``Automatic speech emotion recognition
  using recurrent neural networks with local attention,'' in \emph{IEEE
  International Conference on Acoustics, Speech and Signal Processing
  (ICASSP)}, 2017, pp. 2227--2231.

\bibitem{SER_RNN3}
S.~T. Rajamani, K.~T. Rajamani, A.~Mallol-Ragolta, S.~Liu, and B.~Schuller, ``A
  novel attention-based gated recurrent unit and its efficacy in speech emotion
  recognition,'' in \emph{IEEE International Conference on Acoustics, Speech
  and Signal Processing (ICASSP)}, 2021, pp. 6294--6298.

\bibitem{daiz_2}
S.~H. Dumpala, S.~Rodriguez, S.~Rempel, R.~Uher, and S.~Oore, ``Significance of
  speaker embeddings and temporal context for depression detection,''
  \emph{arXiv preprint arXiv:2107.13969}, 2021.

\bibitem{transformer}
A.~Vaswani, N.~Shazeer, N.~Parmar, J.~Uszkoreit, L.~Jones, A.~N. Gomez,
  L.~Kaiser, and I.~Polosukhin, ``Attention is all you need,'' in
  \emph{Proceedings of the 31st International Conference on Neural Information
  Processing Systems}, 2017, pp. 5998--6008.

\bibitem{ViT}
A.~Dosovitskiy, L.~Beyer, A.~Kolesnikov, D.~Weissenborn, X.~Zhai,
  T.~Unterthiner, M.~Dehghani, M.~Minderer, G.~Heigold, S.~Gelly, J.~Uszkoreit,
  and N.~Houlsby, ``An image is worth 16x16 words: Transformers for image
  recognition at scale,'' in \emph{International Conference on Learning
  Representations}, 2021.

\bibitem{Swin}
Z.~Liu, Y.~Lin, Y.~Cao, H.~Hu, Y.~Wei, Z.~Zhang, S.~Lin, and B.~Guo, ``Swin
  transformer: Hierarchical vision transformer using shifted windows,''
  \emph{International Conference on Computer Vision (ICCV)}, 2021.

\bibitem{speech_use_trans2}
X.~Wang, M.~Wang, W.~Qi, W.~Su, X.~Wang, and H.~Zhou, ``A novel end-to-end
  speech emotion recognition network with stacked transformer layers,'' in
  \emph{IEEE International Conference on Acoustics, Speech and Signal
  Processing (ICASSP)}, 2021, pp. 6289--6293.

\bibitem{meld_c1}
J.~Liang, R.~Li, and Q.~Jin, ``Semi-supervised multi-modal emotion recognition
  with cross-modal distribution matching,'' in \emph{Proceedings of the 28th
  ACM International Conference on Multimedia}, 2020, pp. 2852--2861.

\bibitem{meld_c3}
Z.~Lian, B.~Liu, and J.~Tao, ``Ctnet: Conversational transformer network for
  emotion recognition,'' \emph{IEEE Transactions on Audio, Speech, and Language
  Processing}, vol.~29, pp. 985--1000, 2021.

\bibitem{ksT}
W.~Chen, X.~Xing, X.~Xu, and J.~Yang, ``Key-sparse transformer with cascaded
  cross-attention block for multimodal speech emotion recognition,''
  \emph{arXiv preprint arXiv:2106.11532}, 2021.

\bibitem{IEMOCAP}
C.~Busso, M.~Bulut, C.-C. Lee, A.~Kazemzadeh, E.~Mower, S.~Kim, J.~N. Chang,
  S.~Lee, and S.~S. Narayanan, ``{IEMOCAP: Interactive emotional dyadic motion
  capture database},'' \emph{Language Resources and Evaluation}, vol.~42,
  no.~4, pp. 335--359, 2008.

\bibitem{meld}
S.~Poria, D.~Hazarika, N.~Majumder, G.~Naik, E.~Cambria, and R.~Mihalcea,
  ``{MELD: A} multimodal multi-party dataset for emotion recognition in
  conversations,'' \emph{arXiv preprint arXiv:1810.02508}, 2019.

\bibitem{pitt}
J.~T. Becker, F.~Boiler, O.~L. Lopez, J.~Saxton, and K.~L. McGonigle, ``The
  natural history of alzheimer's disease: Description of study cohort and
  accuracy of diagnosis,'' \emph{Archives of Neurology}, vol.~51, no.~6, pp.
  585--594, 1994.

\bibitem{avec}
J.~Gratch, R.~Artstein, G.~Lucas, G.~Stratou, S.~Scherer, A.~Nazarian, R.~Wood,
  J.~Boberg, D.~DeVault, S.~Marsella, D.~Traum, S.~Rizzo, and L.-P. Morency,
  ``The distress analysis interview corpus of human and computer interviews,''
  in \emph{Proceedings of the Ninth International Conference on Language
  Resources and Evaluation ({LREC}'14)}, May 2014, pp. 3123--3128.

\bibitem{LN}
J.~{Lei Ba}, J.~R. {Kiros}, and G.~E. {Hinton}, ``Layer normalization,''
  \emph{arXiv preprint arXiv:1607.06450}, 2016.

\bibitem{P2FA}
J.~Yuan and M.~Y. Liberman, ``Speaker identification on the scotus corpus,''
  \emph{Journal of the Acoustical Society of America}, vol. 123, pp.
  3878--3878, 2008.

\bibitem{cookie}
K.~E. Goodglass~H, ``The boston diagnostic aphasia examination,'' \emph{Lea \&
  Febinger, Philadelphia}, 1983.

\bibitem{avec2016}
M.~Valstar, J.~Gratch, B.~Schuller, F.~Ringeval, D.~Lalanne, M.~Torres~Torres,
  S.~Scherer, G.~Stratou, R.~Cowie, and M.~Pantic, ``Avec 2016: Depression,
  mood, and emotion recognition workshop and challenge,'' in \emph{Proceedings
  of the 6th International Workshop on Audio/Visual Emotion Challenge}, 2016,
  pp. 3--10.

\bibitem{wav2vec}
S.~Schneider, A.~Baevski, R.~Collobert, and M.~Auli, ``wav2vec: Unsupervised
  pre-training for speech recognition.'' in \emph{Proc. Interspeech}, 2019, pp.
  3465--3469.

\bibitem{sgd}
H.~Robbins and S.~Monro, ``A stochastic approximation method,'' \emph{The
  annals of mathematical statistics}, pp. 400--407, 1951.

\bibitem{iemocap_4}
Y.~Yin, Y.~Gu, L.~Yao, Y.~Zhou, X.~Liang, and H.~Zhang, ``Progressive
  co-teaching for ambiguous speech emotion recognition,'' in \emph{IEEE
  International Conference on Acoustics, Speech and Signal Processing
  (ICASSP)}, 2021, pp. 6264--6268.

\bibitem{pitt_2}
F.~Bertini, D.~Allevi, G.~Lutero, L.~Calzà, and D.~Montesi, ``An automatic
  alzheimer’s disease classifier based on spontaneous spoken english,''
  \emph{Computer Speech \& Language}, vol.~72, p. 101298, 2022.

\end{thebibliography}

\end{document}